\documentstyle[12pt]{article}\textheight 230mm\textwidth 160mm
\newcommand{\bi}{\bibitem}
\newcommand{\be}{\begin{eqnarray}}
\newcommand{\ee}{\end{eqnarray}}

\begin{document}

\catcode`\@=11
\def\lsim{\mathrel{\mathpalette\@versim<}}
\def\gsim{\mathrel{\mathpalette\@versim>}}
\def\@versim#1#2{\vcenter{\offinterlineskip
\ialign{$\m@th#1\hfil##\hfil$\crcr#2\crcr\sim\crcr } }}
\catcode`\@=12

\noindent
\hspace*{11.6cm}\vspace{-2mm}KANAZAWA-95-15\\
\hspace*{11.6cm}August 1995

\begin{center}
{\Large\bf Gauge-Yukawa \vspace{-1mm} Unification $^{\dag}$}
\end{center}

\vspace*{0.1cm}
\begin{center}{\sc Jisuke Kubo}$\ ^{(1)}$,
{\sc Myriam
Mondrag{\' o}n}$\ ^{(2)}$\vspace{-1mm} and \\
{\sc George Zoupanos}$\ ^{(3),*}$
\end{center}
%\vspace*{0.1cm}
\begin{center}
{\em $\ ^{(1)}$  College of Liberal Arts, Kanazawa University,
920-11 Kanazawa, Japan}\\
{\em $\ ^{(2)}$ Institut f{\" u}r Theoretische Physik,
Philosophenweg 16\vspace{-2mm}\\
D-69120 Heidelberg, Germany}\\
{\em $\ ^{(3)}$ Physics Department, National Technical\vspace{-2mm}
University\\ GR-157 80 Zografou, Athens, Greece }
\end{center}

\vspace{1cm}
\begin{center}
{\sc\large Abstract}
\end{center}

\noindent
Gauge-Yukawa Unification (GYU) yields
a functional relation among
the gauge and Yukawa couplings, and it
may follow from the usual Grand Unification
if it is supplemented with some additional
principles.
Postulating the principles of finiteness and
reduction of couplings, we have achieved
 Gauge-Yukawa Unification in various
supersymmetric unified models
leading, among other things, to
interesting predictions on the top quark mass.

\vspace*{4cm}
\footnoterule
\vspace*{5mm}
\noindent
$ ^{\dag}$ Presented by G. Zoupanos at
{\em  SUSY 95, Ecole Polytechnique,
Palaiseau, France, 11-18 May, 1995},
to appear in the proceedings.\\
$^{*}$Partially supported by the C.E.U. projects
(SC1-CT91-0729; CHRX-CT93-0319).

\newpage
\pagestyle{plain}

The traditional way of reducing the number
of the free parameters
of the standard model (SM) is to require that
the theory is more symmetric at higher energies.
This approach has been applied, e.g., in GUTs with
a certain success in the gauge and independently in the Yukawa
sectors of the theory.
However, this attractive principle has its limitation
as it is well known that increasing the gauge symmetry of a GUT
(e.g., $SO(10) ,\, E_6 ,\, E_7 ,\, E_8$ ) one does not
necessarily increase  the predictive power of the theory.
This is because,
 to construct realistic models,
one has to understand the breaking
of these symmetries, which requires introducing
additional free parameters in general.
Alternatively, we suggest \cite{mondragon1}-\cite{kubo3}
that a natural gradual extension
of the GUT idea,
in prospect of increasing the predictability of
the low energy parameters of the theory,
is to attempt to relate the couplings of the gauge and
Yukawa sectors, i.e., to achieve Gauge-Yukawa
Unification (GYU). Searching for a symmetry
that would provide GYU, one is led to consider
$N=2$ supersymmetric theories \cite{fayet},
which  however proved to have more serious phenomenological
problems than the SM.
The last comment holds also for superstring theories and composite
models which could in principle lead
to relations among the gauge and Yukawa couplings.

In our recent studies \cite{mondragon1}-\cite{kubo3},
 we have considered
 the GYU which is based on the
principles of reduction of
couplings \cite{zimmermann1,kubo1,kubo2,kubo3}
 and also finiteness \cite{finite1,finite3,mondragon1}.
These principles, which are formulated in
perturbation  theory, are not explicit symmetry
principles, although they might imply symmetries.
 The former principle is based on the existence
of renormalization group
invariant (RGI) relations among couplings which preserve
 perturbative
renormalizability. Similarly, the latter one is based
on the fact that it is
possible to find RGI
 relations among couplings that
keep finiteness in perturbation theory,
even to all orders \cite{finite3}.
Applying these principles,
one can relate the gauge and Yukawa couplings
 without introducing necessarily a
symmetry, thereby improving the
 predictive power of a model.
In what follows, we briefly outline the basic tool
of this GYU scheme and  its application to various models.

A RGI relation among couplings can be expressed
in an implicit form
\be
\Phi (g_1,\cdots,g_N) ~=~0~,
\ee
which
has to satisfy the partial differential equation (PDE)
$\mu\,d \Phi /d \mu ~=~
\sum_{i=1}^{N}
\,\beta_{i}\,\partial \Phi /\partial g_{i}~=~0$, where
 $\beta_i$ is the $\beta$-function of $g_i$.
There exist ($N-1$) independent  $\Phi$'s, and
finding the complete
set of these solutions is equivalent
to solve the so-called reduction equations \cite{zimmermann1},
\be
\beta_{g} \,\frac{d g_{i}}{d g} &=&\beta_{i}~,~i=1,\cdots,N~,
\ee
where $g$ and $\beta_{g}$ are the primary
coupling and its $\beta$-function,
and $i$ does not include it.
Using all the $(N-1)\,\Phi$'s to impose RGI relations,
one can in principle
express all the couplings in terms of
a single coupling $g$.
 The complete reduction,
which formally preserve perturbative renormalizability,
 can be achieved by
demanding
 power series solution
\be
g_{i} &=& \sum_{n=0}\kappa_{i}^{(n)}\,g^{2n+1}~,
\ee
where the uniqueness of such a power series solution
can be investigated at the one-loop level \cite{zimmermann1}.
The completely reduced theory contains only one
independent coupling with the
corresponding $\beta$-function.
In susy Yang-Mills theories with
a simple gauge group, something more  drastic can happen;
the vanishing of the $\beta$-function
 to all orders in perturbation theory,
if all the one-loop anomalous dimensions of the matter fields
in the completely, uniquely reduced
theory identically vanish \cite{finite3}.

This possibility of coupling unification is
attractive, but  it can be too restrictive and hence
unrealistic. To overcome this problem, one  may use fewer $\Phi$'s
as RGI constraints.
This is the idea of partial
 reduction \cite{kubo1,kubo2,kubo3}, and the power series
solution (3) becomes in this case \be
g_{i} &=& \sum_{n=0}\kappa_{i}^{(n)}(g_{a}/g)
\,g^{2n+1}~,~i=1,\cdots,N'~,~a=N'+1,\cdots,N~.
\ee
The coefficient functions
 $\kappa_{i}^{(n)}$ are required to
be unique power series in $g_{a}/g$ so that the $g_{a}$'s can
be regarded as perturbations to the completely reduced
system in which the $g_{a}$'s identically vanish.
In the following,
 we would like to consider three different GYU models.

\begin{center}
{\bf A. Finite Unified Theory (FUT) based on
 $SU(5)$ \cite{mondragon1}}
\end{center}

This is a $N=1$ susy Yang-Mills theory  based on
 $SU(5)$ \cite{finite2} which contains one ${\bf 24}$,
four pairs of (${\bf 5}+\overline{{\bf 5}}$)-Higgses and
 three
($\overline{{\bf 5}}+{\bf 10}$)'s
for three fermion generations.
The unique power series solution \cite{mondragon1}, which looks
realistic as a first approximation, corresponds to
the Yukawa matrices
without intergenerational mixing, and yields
in the one-loop approximation
\be
g_{t}^{2} &=&g_{c}^{2}~=~g_{u}^{2}~=~\frac{8}{5} g^2~,~
g_{b}^{2} ~=~g_{s}^{2} ~=~g_{d}^{2} ~=~
g_{\tau}^{2} ~=~g_{\mu}^{2} ~=~g_{e}^{2}
{}~=~\frac{6}{5} g^2~,
\ee
where $g_i$'s stand  for the Yukawa couplings.
At first sight, this GYU seems to lead
to unacceptable predictions of the fermion masses.
But this is not the case, because each generation has
an own pair of ($\overline{{\bf 5}}+{\bf 5}$)-Higgses
so that one may assume \cite{finite2,mondragon1}that
after the diagonalization
of the Higgs fields  the effective theory is
exactly MSSM, where the pair of
its Higgs supermultiplets mainly stems from the
(${\bf 5}+\overline{{\bf 5}} $) which
couples to the third fermion generation.
(The Yukawa couplings of the first two generations
can be regarded as free parameters.)
The predictions of $m_t$ and $m_b$ for various $m_{\rm SUSY}$
are given in table 1.

\begin{center}
\begin{tabular}{|c|c|c|c|c|c|}
\hline
$m_{\rm SUSY}$ [GeV]   &$\alpha_{3}(M_Z)$ &
$\tan \beta$  &  $M_{\rm GUT}$ [GeV]
 & $m_{b} $ [GeV]& $m_{t}$ [GeV]
\\ \hline
$200$ & $0.123$  & $53.7$ & $2.25 \times 10^{16}$
 & $5.2$ & $184.0$ \\ \hline
$500$ & $0.118$  & $54.2$ & $1.45 \times 10^{16}$
 & $5.1$ & $184.4$ \\ \hline
\end{tabular}
\end{center}

\begin{center}
{\bf Table 1}. The predictions
for  $m_{\rm SUSY}=200$ and $500$ GeV for FUT.
\end{center}

\begin {center}
{\bf B. Asymptotically free
Dimopoulos-Georgi-Sakai (DGS) Model \cite{kubo2} }
\end{center}

The field content is minimal. Neglecting the
Cabibbo-Kobayashi-Maskawa mixing, there are six
Yukawa and two Higgs
couplings at the beginning. We then require
GYU to occur among the Yukawa couplings of the third
generation and the gauge coupling.
We also require the theory to
be completely asymptotically free.
In the one-loop approximation, the GYU yields
$g_{t,b}^{2} ~=~\sum_{m,n=1}^{\infty}
\kappa^{(m,n)}_{t,b}~h^m\,f^n~g^2$.
($h$ and $f$ are related
to the Higgs couplings.)
$h$ is allowed to vary from
$0$ to $15/7$, while $f$ may vary from $0$ to a maximum which depends
on $h$ and vanishes at $h=15/7$.
As a result, we obtain\cite{kubo2}
\be
0.97\,g^2 \lsim g_{t}^{2} \lsim 1.37\,g^2~,
{}~0.57\,g^2 \lsim g_{b}^{2}=g_{\tau}^{2}  \lsim 0.97\,g^2~.
\ee
In table 2, we give some representative  predictions.

\begin{center}
\begin{tabular}{|c|c|c|c|c|c|c|c|}
\hline
$m_{\rm SUSY}$ [GeV]
& $g_{t}^{2}/g^2$ & $g_{b}^{2}/g^2$&
$\alpha_{3}(M_Z)$ &
$\tan \beta$  &  $M_{\rm GUT}$ [GeV]
 & $m_{b} $ [GeV]& $m_{t}$ [GeV]
\\ \hline
$300$ & $1.37$ & $0.97$ & $0.120$  & $52.2$ & $1.9\times 10^{16}$
  & $5.2$  & $182.8$ \\ \hline
$300$ & $0.97$& $0.57$ & $0.120$  & $47.7$ & $1.8\times10^{16}$
  & $5.4$  & $179.7 $  \\ \hline
$500$ & $1.37$ & $0.97$ & $0.118$  & $52.4$  & $1.43\times10^{16}$
  & $5.1$  & $182.7$ \\ \hline
$500$ & $0.97$& $0.57$ & $0.118$  & $47.7$ & $1.39\times10^{16}$
  & $5.3$  & $178.9$  \\ \hline
\end{tabular}
\end{center}

\begin{center}
{\bf Table 2}. The predictions
of the asymptotically free DGS model
\end{center}

\begin{center}
{\bf C. Asymptotically non-free SUSY
Pati-Salam (PS) Model \cite{kubo3}}
\end{center}

This is a model
 without covering GUT \cite{pati1}, that is, there is no
gauge coupling unification as it stands.
The field content is \cite{kubo3}:
three
$({\bf 4},{\bf 2},{\bf 1})$ and three
$({\bf \overline{4}},{\bf 1},{\bf 2})$
under $SU(4)\times SU(2)_{\rm L}\times SU(2)_{\rm R}$
for three fermion generations,
a set of $({\bf 4},{\bf 2},{\bf 1})$,
$({\bf \overline{4}},{\bf 2},{\bf 1})$
and  two $({\bf 15},{\bf 1},{\bf 1})$ for
 Higgses that are responsible for the spontaneous
symmetry breaking down
to $SU(3)_{\rm C}\times SU(2)_{\rm L}\times U(1)_{Y}$,
and also a set of
 $({\bf 1},{\bf 2},{\bf 2})$,
 $({\bf 15},{\bf 2},{\bf 2})$ and
$({\bf 1},{\bf 1},{\bf 1})$.
The singlet supermultiplet mixes with the right-handed neutrino
supermultiplets at a high energy scale, while
$({\bf 15},{\bf 2},{\bf 2})$ is introduced to realize
the  Georgi-Jarlskog type ansatz for
the fermion mass matrix.

In one-loop order,  we first obtain the unification of the gauge
couplings,
\be
g_{4}^{2} &=& \frac{8}{9}\,g_{2\rm L}^{2}~,~
g_{2\rm R}^{2} ~=~ \frac{4}{5}\,g_{2\rm L}^{2}~.
\ee
 In the Yukawa sector,
we find
\be
2.8\,g_{2\rm L}^{2}
 \lsim g^{2}_{t}~=~g^{2}_{b}~=~g^{2}_{\tau} \lsim 3.5
\,g_{2\rm L}^{2}~. \ee
The typical predictions are presented in table 3.

\begin{center}
\begin{tabular}{|c|c|c|c|c|c|c|c|}
\hline
$m_{\rm SUSY}$ [GeV]
& $g_{t}^{2}/g_{2 \rm L}^{2}$ &
$\alpha_{3}(M_Z)$ &
$\tan \beta$  &  $M_{\rm GUT}$ [GeV]
 & $m_{b} $ [GeV]& $m_{t}$ [GeV]
\\ \hline
$500$ & $2.8$ & $0.129$  & $61.2$ & $0.16\times 10^{16}$
  & $5.4$  & $196.8$ \\ \hline
$500$ & $3.4$& $0.132$ & $62.1$ & $0.17\times 10^{16}$
  & $5.4$  & $198.3$ \\ \hline
$1600$ & $2.8$& $0.114$   & $62.5$ & $0.07 \times 10^{16}$
  & $4.8$  & $192.7$  \\ \hline
$1600$ & $3.4$ & $0.112$   & $63.4$ & $0.06\times 10^{16}$
  & $4.7$  & $193.3$ \\ \hline
\end{tabular}
\end{center}
\begin{center}
{\bf Table 3}. The predictions
of the asymptotically non-free SUSY Pati-Salam model
\end{center}

In all of the analyses above,
 we have used the
RG technique and regarded the GYU relations
(5)-(8) as
 the boundary conditions  holding
at the unification scale $M_{\rm GUT}$.
We have assumed that
it is possible to arrange
the susy mass parameters along with
the soft breaking terms in such a way that
the desired symmetry breaking pattern
 really occurs, all the superpartners are
unobservable at present energies,
there is no contradiction with proton decay,
and so forth.
To simplify our numerical analysis  we have also assumed a
unique  threshold $m_{\rm SUSY}$ for all
the  superpartners.
Then we have examined numerically the evolution of the gauge
and Yukawa couplings below $M_{\rm GUT}$ including the two-loop
 effects.

Concerning recent related studies by other authors,
we would like to emphasize that
our approach of dealing with asymptotically non-free
theories \cite{kubo3} covers ref. \cite{ross}
though the underlying idea might be different.
In ref. \cite{jack}, interesting RGI relations among
the soft breaking parameters above
the unification scale has been found.
These relations are obtained on the
close analogy of our approach presented here.

\end{document}